\documentclass[%
 reprint,
 amsmath,amssymb,
 aps,nofootinbib,   
]{revtex4-1}

\usepackage{amsmath}
\usepackage{bm}
\PassOptionsToPackage{linktocpage}{hyperref}
\usepackage[hyperindex,breaklinks]{hyperref}
\usepackage{enumitem}
\usepackage{slashed}

\usepackage{xcolor}
\usepackage{float}

 \usepackage{bbold} 
\newcommand{\de}{\mathrm{d}}	

\usepackage{array}
\usepackage{mathtools}

\usepackage{etoolbox}

\begin{document} 

\title{Complexity and Time}

\author{  C\'esar G\'omez} 
\affiliation{Instituto de F\'{i}sica Te\'orica UAM-CSIC, Universidad Aut\'onoma de Madrid, Cantoblanco, 28049 Madrid, Spain}


\begin{abstract}
For any quantum algorithm given by a path in the space of unitary operators we define the computational complexity as the typical computational time associated with the path. This time is defined using a quantum time estimator associated with the path. This quantum time estimator is fully characterized by the Lyapunov generator of the path and the corresponding quantum Fisher information. The computational metric associated with this definition of computational complexity leads to a natural characterization of cost factors on the Lie algebra generators. Operator complexity growth in time is analyzed from this perspective leading to a simple characterization of Lyapunov exponent in case of chaotic Hamiltonians. The connection between complexity and entropy is expressed using the relation between quantum Fisher information about quantum time estimation and von Neumann entropy. This relation suggest a natural bound on computational complexity that generalizes the standard time energy quantum uncertainty. The connection between Lyapunov and modular Hamiltonian is briefly discussed. In the case of theories with holographic duals and for those reduced density matrix defined by tracing over a bounded region of the bulk, quantum estimation theory is crucial to estimate quantum mechanically the geometry of the tracing region. It is suggested that the corresponding quantum Fisher information associated with this estimation problem is at the root of the holographic bulk geometry.

 \end{abstract}
\maketitle


\section{Introduction and summary}
A concept that has received recently a lot of attention in the general framework of holography as well as in quantum information is the notion of computational complexity (see for instance \cite{Susskind1,Susskind2, Susskind3,Susskind4,Susskind5,Susskind6, Myers1,Myers2,Aaronson,Bala}). In a nutshell the computational complexity of a quantum computer  program  measures the number of elementary gates we need to use to define the program. In this sense we can assign to any quantum algorithm its computational complexity. Generically a quantum algorithm defines a path $|\psi(s)\rangle$ in the Hilbert space of the quantum computer where the initial state $|\psi(s=0)\rangle$ encodes the available data and the final state $|\psi(s=1)\rangle$ provides, after measuring and with high enough probability, the answer to the problem the quantum algorithm is trying to solve. The program itself is associated with the particular path. 

In quantum mechanics the Hilbert space of states is equipped with a natural notion of geometry that measures the distinguishability of states and can be defined for pure states as well as for mixed states \cite{W,FS1,FS2,FS3,aho}. This distinguishability metric naturally associates with the path defining the quantum algorithm a length and a natural question is what is the relation between this length and the computational complexity of the program. This question was first addressed, for the particular case of Grover algorithm \cite{Grover}, in \cite{Gomez1}. In this case the path in Hilbert space defined by the quantum algorithm is a geodesic in the Fubini Studi metric.

A complementary view of the quantum algorithm can be defined using paths in the space of unitary transformations. Given the quantum algorithm path $|\psi(s)\rangle$ we can define an operator $U$ such that $U|\psi(0)\rangle = |\psi(1)\rangle$ and to define the program as a path in the space of unitary operators $O(s)$ such that $O(s=0)=1$ and $O(s=1) =U$. In this case the computational complexity of a given program is defined as the number of gates we need to compose to define such a path. 

An interesting possibility is to define in the space of unitary operators a notion of distance $d(1,U)$ that measures the minimal number of quantum gates we need to compose in order to get $U$ from the identity. This distance that has received the name of relative complexity \cite{Susskind5} depends on how we select the subspace of unitary operators that qualify as elementary gates as well as on how we characterize the error we are able to accept i.e. the minimal distance between the result of the gate composition and the desired operator $U$. 

A natural possibility is of course to define the relative complexity distance using a computational metric $G^c_{I,J}(O)$ in the space of unitary operators in such a way that the length of the geodesic going from $1$ to the operator $U$ coincides with $d(1,U)$. An interesting attempt in this direction is to modify the natural metric on the space of unitary operators using the so called cost functions $c(I,O)$ that effectively pushes, at each point, the geodesic to be generated by a vector field living in the subspace of the Lie algebra generated by the elementary gate operators \cite{Nielsen1,Nielsen2,Nielsen3} (see also \cite{Myers1,Bala}).

A different but intimately related issue is the complexity of time evolution and the operator complexity growth with time \cite{operator1,operator2,operator3}. In this case the path is generated by a unique unitary operator $H$ that we can identify with the Hamiltonian. Thus $O(t)= e^{-iHt}O(0)e^{iHt}$ and analogously for $|\psi(t)\rangle$. In the case $[H,O(0)] \neq 0$ we are interested in how a would be simple initial operator $O(0)$ becomes with time more and more complex. This problem requires to decompose the space of unitary operators in different subspaces with different degree of complexity where that, for a lattice system, can be defined as the number of sites where the operator is acting non trivially. Other decompositions of the space of unitary operators can be used to measure how the initial operator living in one subspace spreads over time on other subspaces. 

In this note we shall take a slightly different approach to the notion of computational complexity. For the concrete case of a program defined by a path $O(s)$ going from the identity to some given operator $U$ we will like to estimate the complexity of the program as {\it the typical physical time} the computer needs to develop this program. Since the program is defined using a formal dimensionless parameter $s$ that by construction goes from $0$ to $1$ to estimate the time requires to define some {\it time estimator} $T(s;O(s))$ associated with the path $O(s)$ in such a way that a measure of the time needed to develop the program could be formally defined as
\begin{equation}
C(O(s)) = \int_0^1 \de s T(s;O(s))\,.
\end{equation}
The proposal suggested in this note is to use in order to define $T$ and consequently the complexity $C$ the theory of quantum estimation \cite{QE1,QE2,QE3,QE4,QE5,QE6}. Given the path $O(s)$ we can define a self adjoint operator $\hat S(s)$ that effectively estimate the value of the parameter $s$. This parameter is some sort of external computer time that goes from zero to one according with the way the computer is developing the program. However this external computer time is not telling us in reality how much physical time the computer is really taking. A measure of this physical time can be defined using the intrinsic quantum uncertainty we have to estimate the external time $s$ at which the computer is working using as data the set of measurements that we can do over the state of the quantum computer. This uncertainty is measured by the standard deviation $\Delta(\hat S^2)(s)$ of the quantum estimator. This is the quantity we shall use to define $T(s;O(s))$, namely
\begin{equation}
T(s;O(s)) = \sqrt{\Delta(\hat S^2)(s)}\,,
\end{equation}
and consequently also to measure the complexity of a given path $O(s)$
\begin{equation}
C(O(s)) = \int_0^1 \de s \sqrt{\Delta(\hat S^2)(s)}\,.
\end{equation}
The technical steps in the definition of the {\it time quantum estimator} are roughly the following. Given the path $O(s)$ we first define the Lyapunov operator $L_s$ by solving the Lyapunov equation
\begin{equation}
\frac{L_sO(s) +O(s)L_s}{2} = \frac{\de O(s)}{\de s}\,.
\end{equation}
Once we have found $L_s$ we define the {\it quantum Fisher information function} $F(s)$ by
\begin{equation}
F(s) = Tr[O(s) L_s^2]
\end{equation}
and the time quantum estimator $\hat S(s)$ as
\begin{equation}\label{estimator}
\hat S(s) = s \mathbb{1} + \frac{L_s}{F(s)}\,.
\end{equation}
For the operator $\hat S(s)$ the Cramer-Rao theorem \cite{Cramer1, Cramer2} allows us to discover a bound on $\Delta(\hat S^2)(s)$ and to define the corresponding complexity. 

The natural relation between the quantum Fisher function and the distinguishability metric on the Hilbert space of states allows us to have a very intuitive geometrical interpretation of the complexity defined above. It is simply the length of the path in the space of unitary operators but relative to a metric that is defined by the inverse of the quantum Fisher function. In essence this complexity metric is the {\it dual} to the standard distinguishability metric, where in essence this duality simply reflects {\it the time energy uncertainty}.

Before going into technicalities it will be worth, in order to set the frame of the discussion, to make some general philosophical remarks. From a physics point of view the essential target of the discussion lies in understanding the differences as well as the connections between computational complexity and entropy. Generically for a given system complexity can be much larger than entropy and therefore the system can evolve increasing complexity even after reaching its maximal entropy. This observation, as stressed in \cite{Susskind2}, can be specially important in the context of black hole physics.  Our approach to complexity based on time estimation is related to quantum Fisher information by contrast to the von Neuman or quantum Shanon information underlying the notion of entropy as well as that of entanglement. The key difference is that quantum Fisher information depends on the time variation of the system while entropy is an equilibrium notion independent on how the probabilities change with time. Thus, the relation between complexity and entropy, from that point of view, should reflect the relation between the information about the state  (entropy) and that about the time (complexity). This relation reflects one of the most basics principles of quantum mechanics, namely, the time energy uncertainty relation that, as we shall discuss at the end of this note, leads to a qualitative relation between complexity ${\cal{C}}$ and entropy $S$ of the type
\begin{equation}
{\cal{C}} \leq e^{S}\,.
\end{equation}
The former relation is a consequence of Cramer-Rao theorem, thus if we define the complexity using an optimal quantum time estimator we can saturate the inequality. In this case the complexity $e^S$ associates a time scale to a given entropy $S$.
In what follows we shall try to make these general comments as precise as possible.

\section{Lyapunov equation and quantum Fisher function}
Let us consider a path of operators $O(s)$ and let us define a one parameter family of eigenbasis $|\phi_n(s)\rangle$ by
\begin{equation}
O(s) |\phi_n(s)\rangle = \rho_n(s) |\phi_n(s)\rangle\,.
\end{equation}
In this basis we have $O(s)= \sum_n \rho_n(s) |\phi_n(s)\rangle \langle \phi_n(s)|$. 
Let us now consider the Lyapunov equation for $O(s)$ and solve it discovering $L_s$ \cite{QE3}. What we get is simply
\begin{align}
L_s =& \sum_n \left( \frac{\partial_s \rho_n(s)}{\rho_n(s)} |\phi_n(s)\rangle \langle \phi_n(s)| \nonumber \right. \\
&+ 2\rho_n(s) \sum_{m\neq n} L_{n,m}(s)|\phi_n(s)\rangle \langle \phi_m(s)| \nonumber \\
&\left. +2\rho_n(s) \sum_{m\neq n} L_{n,m}(s)|\phi_m(s)\rangle \langle \phi_n(s)|\right)\,,
\end{align}

where $|\partial_s \phi_n(s)\rangle = \sum_m L_{m,n} |\phi_m(s) \rangle$.

For future use is interesting to find the Lyapunov operator $L_s$ in the case of Hamiltonian evolution \cite{QE3}. In this case $O(s) = e^{-iHs}O(0)e^{iHs}$ for a given $H$. Using for $O(0)$ the representation $O(0)= \sum_n \rho_n |\phi_n\rangle\langle\phi_n|$ we get $L_s=e^{iHs}L_0e^{-iHs}$ with
\begin{equation}\label{one}
L_0 = 2i \sum_{n,m} \frac{\langle \phi_m|[H,O(0)]|\phi_n\rangle}{\rho_m+\rho_n} |\phi_n\rangle \langle \phi_m|\,.
\end{equation}
Once we have defined the Lyapunov operator $L_s$ we define the quantum Fisher function
\begin{equation}
F(s) = Tr[O(s) L_s^2]\,.
\end{equation}
At this point is easy to check that
\begin{equation}
Tr[O(s)L_s] =0\,,
\end{equation}
and that in the case of hamiltonian time evolution the quantum Fisher function is constant and given by
\begin{equation}
F= Tr [O(0)L_0^2]\,.
\end{equation}
\section{The Fubini Studi metric}
Let us  consider the quantum computer Hilbert space ${\cal{H}}$ of $n$ q-bits and dimension $N=2^n$. The coordinates of a state are $2^n$ complex numbers $c_i$. Introducing phases $c_i=\sqrt{p_i}e^{\phi_i}$ we can write the FS metric as
 
 \begin{equation}
 \de s^2_{FS} = \frac{1}{4}\sum \frac{\de p_i^2}{p_i} +\left(\sum p_i\de \phi_i^2 -\left(\sum p_i\de \phi_i\right)^2\right)\,.
 \end{equation}
 
 Any path $|\psi(s)\rangle$ of pure states is defined by $p_i(s),\phi_i(s)$ and the induced metric on the path is given by
 \begin{equation}
G \equiv \left(\frac{1}{4}{\cal{F}}(s) +  \sigma(s)^2\right) \de s^2 \,,
 \end{equation}
 where ${\cal{F}}$ is the {\it classical} Fisher information function
 \begin{equation}\label{Fisher}
 {\cal{F}} = \sum \frac{\dot p_i^2}{p_i}
 \end{equation}
 and $\sigma$ is the standard deviation
 \begin{equation}
 \sigma = \sqrt{\sum p_i x_i^2 -(\sum x_ip_i)^2}
 \end{equation}
 for a set of random variables $x_i\equiv \dot \phi_i$. It is easy to see that this induced metric is just defined by the quantum Fisher information as 
 \begin{equation}
 G= 
 F(s)\de s^2\,.
 \end{equation}
 In the simple case of pure states we have $O(s)\equiv |\psi(s)\rangle \langle \psi(s)|$ and $F(s) = Tr[O(s) \de O(s)^2]$.\footnote{Representing the quantum state in terms of $p_i$ and $\phi_i$ as before we get for the real part of the quantum Fisher information metric
 \begin{equation}
 G_q= \frac{1}{4}E((\de \ln p)^2) +E((\de \phi)^2)-(E(\de \phi))^2\,,
 \end{equation}
 where $E$ represents the average relative to $p_i$ i.e $E(f) = \sum p_i f(i)$.This metric is the FS metric. The imaginary part is a symplectic form given by $-iE(\de \ln p \wedge \de \phi)$ and defines a Berry phase (see \cite{Marmo}). } The same can be done for mixed states using the corresponding quantum Fisher function. This metric is known as Bures metric \cite{FS3}.
 \section{The quantum length of time evolution and a new view on chaos bounds}
 This is an especially simple case when we consider pure states. For the time evolution path $|\psi(t)\rangle$ defined by a Hamiltonian $H$ the length of the path for a given time $t$ is simply
 \begin{equation}\label{length}
 L(|\psi(0)\rangle; H,t) = t\sqrt{F}\,,
 \end{equation}
 with $F=Tr[O(0)L_0^2]$ where $O(0)=|\psi(0)\rangle\langle \psi(0)|$ and $L_0$ given in (\ref{one}). This is nothing else but the standard deviation $\Delta(H^2)$ of the energy in the initial state $|\psi(0)\rangle$. The meaning of this length is quite transparent. It is just the total time of the Hamiltonian evolution measured in a time unit defined by $\frac{1}{\sqrt{F}}$ that simply defines the distinguishability time for this Hamiltonian for the state $|\psi(0)\rangle$.\footnote{Note that (\ref{length}) simply reflects the energy time uncertainty relation in the sense originally described in \cite{Mandelstam}. For a discussion on this issue see \cite{aho,Fock,BA}.} This quantity can be interpreted as a measure of the complexity of the time evolution. For fixed time $t$ this complexity becomes maximal for the state for which $\sqrt{F}$ has its maximal value.\footnote{This is the value of $\Delta(E)$ on the corresponding state, so it can be bounded once we couple the system to gravity.} 
 
 More interesting is the case of operator time evolution $O(t)$ for some initial operator $O(0)=O$. In this case we are interested in the correlation function
 \begin{equation}
 C(t) = Tr[OO(t)]\,.
 \end{equation}
 In \cite{operator3} the growth of complexity with time of the operator, the so called K-complexity, is defined introducing the Krylov basis of operators associated with $O$ ( see also \cite{Barbon}). The moments of $C(t)$  are introduced as 
\begin{equation}
 \mu_{2n} = \left.\frac{\de ^nC(t)}{\de t^n} \right|_{t=0}\,.
 \end{equation}
 
 Using the corresponding Lyapunov operator we have
 \begin{equation}\label{two}
 \mu_2= \frac{\de^2 C(t)}{\de t^2}|_{t=0}= Tr[O L_0^2]\,, 
 \end{equation}
 which is nothing else but the quantum Fisher function $F(O;H)$ that depends on $H$ and the operator $O$. Using the eigenbasis of the operator $O$ we get \cite{QE3}
 \begin{equation}
\mu_2 = 2\sum_{m\neq n} \sigma_{m,n} H_{m,n}^2\,, 
\end{equation}
with $\sigma_{m,n} = \frac{(\rho_m-\rho_n)^2}{\rho_n+\rho_m}$ plus any antisymmetric term.
 
For chaotic Hamiltonians we can use the ETH \cite{Srednicki1,Srednicki2}  that implies that the eigen basis of $O$ and that of the Hamiltonian $H$ are uncorrelated \cite{Barbon}. In this case we can approximate
\begin{equation}
\mu_2 \sim Tr[O\Delta H^2]\,,
\end{equation}
which is the energy uncertainty in the density matrix $\sum_n \rho_n|\phi_n\rangle\langle\phi_n|$ defined by $O$. 
We can use this fact to estimate the corresponding Lyapunov exponent as
\begin{equation}
\lambda \sim \sqrt{Tr[O\Delta H^2]}\,,
\end{equation}
where we have taken into account the normalization factor $O(e^n)$ for $n$ the number of degrees of freedom. This estimate of the Lyapunov exponent naturally leads to a bound as 
 \begin{equation}\label{bound}
 \lambda \sim \frac{T}{\hbar}\,,
 \end{equation}
 for $T$ the effective temperature in the thermalization limit. In summary if the system is chaotic meaning totally uncorrelated eigenbasis of $O$ and $H$ we conjecture that the corresponding Lyapunov exponent is bounded by the standard deviation of energy $\sqrt{Tr[O\Delta H^2]}$. This is in agreement with the bound on chaos in \cite{Malda}. It could be instructive to compare this  discussion with the one for the SYK model where you get \cite{operator2}, in agreement with (\ref{bound}),
 \begin{equation}
 \mu_2 \sim {\cal{J}}^2
 \end{equation}
 for ${\cal{J}}$ the SYK energy scale.
 \section{Space of unitary operators and information cost}
 Let us now consider the space of unitary transformations $U(N)$. A nice way to think this space is as the space ${\cal{B}}$ of orthonormal basis. Now we can visualize ${\cal{B}}$ as a fiber bundle. Let us consider the equivalence class of orthonormal basis that have the same first element. The space of equivalence classes is isomorphic to $\frac{U(N)}{U(N-1)}$ which is itself isomorphic to the set $S_{N-1}$ of unit vectors in ${\cal{H}}$. So we get the bundle
\begin{equation}\label{two}
 U(N) \rightarrow S_{N-1} \rightarrow U(N-1)\,,
\end{equation}
with projection $\Phi:U(N) \rightarrow S_{N-1}$. Moreover we can think $S_{N-1}$ as a bundle with base space the complex projective space $CP(N-1)$ and fiber $U(1)$ representing the standard phase in quantum mechanics. Combining both we get the space of unitary transformations $U(N)$ as a bundle on $CP(N-1)$ with fiber $U(N-1)\times U(1)$
\begin{equation}
U(N) \rightarrow CP(N-1) \rightarrow U(N-1)\times U(1)\,.
\end{equation}
Let us consider the bundle (\ref{one}). The natural metric on $U(N)$ is given by
\begin{equation}\label{metric}
G(X,Y) = 2Tr(XY)\,,
\end{equation}
for $X,Y$ in the tangent bundle of $U(N)$. Note that this metric when reduced to $SU(N)$ is the standard Cartan Killing metric. The projection $\Phi$ is relative to this metric a Riemannian submersion. This means that the projection defines an isometry between the tangent space of the base $TS_{N-1}$ and the horizontal part of the tangent space of $U(N)$. In other words, this means that {\it horizontal geodesics in the space $U(N)$ project into geodesics in $S_{N-1}$}. 

Let us denote $Y_{I}$ the generators of the Lie algebra and let us consider a given point $O$ in the space of unitary operators. We can define evolutions $O_{\lambda_I}$ generated by the corresponding element in the Lie algebra $Y_{I}$. For each Lie algebra generator we can define the corresponding Lyapunov generator
$L_{\lambda_{I}} \equiv L_{I}$ at the point $O$ in the usual way. This allows us to define the quantum Fisher metric at this point in the space of unitary transformations as
\begin{equation}\label{metric}
F(O)_{I,J} = Tr[\partial_{I }O L_{J}]\,.
\end{equation}
This definition of the metric leads to a natural definition of costs. Imagine you are at the point $O$. The program of the quantum computer needs to decide with what element in the Lie algebra to act. A cost associated with the direction $I$ in tangent space can be defined as
\begin{equation}\label{costs}
c(I,O) = \frac{1}{F(O)_{II}}\,,
\end{equation}
i.e. as the inverse of the quantum Fisher information in $I$ direction. The logic for this definition of costs will be explained in a moment.
\section{Quantum estimator and Complexity}
Once we have introduced the main technical tools we can move into the definition of complexity using the quantum time estimator. In the simplest case of just one parameter the optimal quantum estimator is defined by (\ref{estimator}). The Cramer-Rao theorem is telling us that 
\begin{equation}
\Delta (\hat S^2)(s) \equiv Tr(O(s) \hat S(s)^2) - s^2 = \frac{1}{F(O,s)}
\end{equation}
and consequently we define the complexity of the path $O(s)$ as
\begin{equation}
C(O(s)) = \int _{0}^{1} \sqrt{\Delta (\hat S^2)(s)}\,.
\end{equation}
To see the geometrical meaning of this expression let us parametrize the path $O(s)$ using coordinates $\lambda_{I}(s)$ with $I$ running on the tangent space of unitary operators. Now we can use the metric (\ref{metric}). Using the generalization of Cramer-Rao theorem we define
\begin{equation}\label{complexity}
C(O(s)) = \int_{0}^{1} \de s \sqrt{ F(O)_{IJ}^{-1} \partial_s \lambda_{ I} \partial_{s} \lambda_{J}}\,.
\end{equation}
In other words we define the complexity metric as $F(O)_{IJ}^{-1}$ that roughly correspond to use as cost factors on the Lie algebra the ones defined in (\ref{costs}).

Now we can consider the space of paths $O(s)$ going from the identity into some fixed operator $U$ and to minimize on the space of paths the former definition of complexity. The costs we are introducing through the quantum Fisher metric have a very natural meaning. The computer should choose directions in the Lie algebra that are the most efficient in changing the state of the quantum computer. 

\section{Costs, quantum error correction and locality}
Any quantum computer program should be supplemented by a quantum error correcting code $QEC$ (see for instance \cite{Gotes}). If we think the program as defined by a path of Hamiltonians $H(s)$ we should impose that whenever $H(s)$ acts on the state $|\psi(s)\rangle$ of the quantum computer  the {\it errors} induced by the action of $H(s)$ should be errors that can be corrected. To be more precise at any moment $s$ of the computation we must assume that if the quantum state of the computer is in the code subspace the action of the corresponding Hamiltonian gate $H(s)$ should move the state, up to errors that can be corrected, into the code subspace. This gives us a natural characterization of how to introduce costs, namely we should penalize those gates that induce errors that cannot be corrected. 

Normally costs are introduced on the basis of a locality characterization of gates. Simple gates are those acting locally on a reduced number of q-bits. This is a definition that is very dependent on what basis we choose. We, by contrast, suggest to define simplicity of gates in a way relative to the QEC code implemented in the computer. 

This also gives us a notion of locality. The distance $d$ of the QEC code sets the maximal number of q-bits where we can commit an error that can be corrected by the QEC code. The intrinsic errors of the program can be characterized by the standard deviations $\Delta (\lambda_I)$. Let us define a modified Hamiltonian by $H+\Delta(H)$ where
\begin{equation}
\Delta(H) \equiv \sum_{I} \Delta(\lambda_I) Y^{I}\,.
\end{equation}
In order to define a safe program what we need is that the action of $\Delta(H)$ on a generic state in the code subspace gives rise to errors that can be corrected. This induces a QEC code notion of {\it locality}, namely $\Delta (H)$, must act on a maximum of $d$ q-bits, for $d$ the distance of the code.

As discussed in \cite{gomez2} if the quantum program satisfies this condition the path of Hamiltonians $H(s)$ defining the program are related by RG transformations. Formally
\begin{equation}
H(\alpha s) = T_{\alpha} H(s)\,,
\end{equation}
for $T_{\alpha}$ the corresponding RG transformation. In this case the complexity is fully determined by the formal beta functions $\beta_{I} = \frac{\partial \lambda_I}{\partial s}$ contracted with the inverse of the quantum Fisher metric $F(O)_{IJ}$.

\section{Final Comments/Conjectures}
\subsection{Lyapunov versus modular: holography}
The Lyapunov generator $L_{\rho}$ that we have discussed is a typical state dependent operator associated with a one parameter family of density matrix. In axiomatic quantum field theory a state dependent operator is the modular Hamiltonian \cite{Haag}, \cite{Myersmod} defined by
\begin{equation}
 H_m \equiv-\ln\rho\,.
\end{equation}
 Let us consider the reduced density matrix associated with a certain region $A$. This reduced density matrix depends on the parameters defining the region $A$ over which we are tracing. The standard deviation $\Delta H^2_m$ of the modular hamiltonian is intimately related with the entanglement von Neumann entropy. We can also consider the Lyapunov Hamiltonian associated with small modifications of the region $A$. The corresponding standard deviation is now related with the quantum Fisher information that measures our ability to estimate the size of $A$. Thus we have two relevant quantities one is the entanglement entropy that is telling us how much information we are missing ignoring what happens inside region $A$ {\it but} also the information we need in order to estimate, performing quantum measurements, the geometrical shape of $A$. Both quantities are again related in a way similar to the energy time uncertainty but now relating interior of $A$ and shape of $A$. 
 
 In order to make contact with holography, a notion that we have not used in our former discussion, we can consider the following simple example. Imagine the reduced density matrix defined by tracing on a spherical region $A_r$ of size $r$. Now let us ask ourselves how {\it to estimate the value of this size} $r$ on the basis of the knowledge of measurements performed on $\rho_{A_r}$. How good can be this estimate is encoded in the corresponding quantum Fisher metric. More specifically in the component of the quantum Fisher metric
 \begin{equation}
 G_{rr}(\rho_{A_r})\,,
 \end{equation}
 where we make explicit the dependence on the concrete operator $\rho_{A_r}$. We shall conjecture that this quantum metric is the root of the holographic metric. For instance think in $A_r$ as a spherical region in the bulk with $r$ the bulk coordinate. The holographic metric $g_{rr}$ should be fully determined by the quantum Fisher metric $G_{rr}$. 
 
 There are some hints in this direction. In \cite{taka1,taka2} the relation between quantum Fisher information and holographic metric has been recently suggested for CFT's.\footnote{The holographic meaning of Fisher metric was also discussed in \cite{vR}.} The suggested connection between GR and QM \cite{SusskindGR, Malda2} can be probably rephrased saying that GR emerges as the statistical quantum metric of QM. In that sense entanglement i.e. von Neumann is not enough \cite{Susskind2}; you need Fisher that accounts for the quantum information we have about the shape of the tracing region. Very roughly the space time metric $g_{ss}$ in the estimator $s$ direction should be related with the corresponding quantum Fisher metric $g^q$ as
 \begin{equation}
 \frac{1}{\sqrt{g_{ss}}} \sim \sqrt{g^q} +1\,,
 \end{equation}
 reflecting, very qualitatively, the emergence of space-time metric from quantum statistics.
 
 \subsection{Complexity-entropy duality}
 There exist a well known relation between Fisher information $F(p)$ and Shanon information $S(p)$  (see \cite{Stam} for details) given by:
 \begin{equation}\label{Stam}
 F(p) \geq \frac{1}{e^{2S(p)}}\,.
 \end{equation}
 This relation is a sort of generalized quantum uncertainty relation and is essentially equivalent to the Cramer-Rao theorem \cite{Stam}.
 In this note we have suggested a definition of complexity that essentially goes like ${\cal{C}} \sim \frac{1}{F}$. This leads to the qualitative relation we have advertised in the introduction, namely
 
\begin{align}
\mathcal{C} \leq e^{S}\,,
\end{align}

with the inequality saturated in the case of using optimal estimators.
This relation explains why complexity is exponentially larger than entropy. Notice that (\ref{Stam}) can be written in a nice way as $F e^S \geq 1$
where $F$ is information needed to estimate time and $e^S$ can be interpreted, in micro canonical sense, as the total number of states with the same energy. Heuristically we can think $e^S$ as the number of states degenerate in energy and the complexity as the time needed to distinguish quantum mechanically these states, a time that will go as $\frac{1}{\Delta E} e^S$. We would like to suggest that this is a universal relation telling us that systems with finite entropy cannot be eternal in the sense they have associated a natural upper time scale, namely the time needed to distinguish quantum mechanically the states contributing to the entropy. This result has a natural cosmological reading for the case of de Sitter.\footnote {This cosmological comment that is beyond the scope of this note and will be discussed elsewhere is, for instance, related with the claims about the finite life of de Sitter discussed in \cite{us}. }
 
{\bf Acknowledgements.}
 This work was supported  by   the grants SEV-2016-0597, FPA2015-65480-P and PGC2018-095976-B-C21. 

\end{document}